%% file: LHCP2014.tex
\documentclass[10pt]{article}
\usepackage{graphicx}
\usepackage{amsmath}
\newcommand{\unitm}[1]{\,{\rm{#1}}}

\def\Title#1{\begin{center} {\Large #1 } \end{center}}
\def\Author#1{\begin{center}{ \sc #1} \end{center}}
\def\Address#1{\begin{center}{ \it #1} \end{center}}

\newcommand\pubblock{\rightline{\begin{tabular}{l} Proceedings of the Second Annual LHCP\\ 
         \pubdate  \end{tabular}}}
\newenvironment{Abstract}{\begin{quotation} \begin{center} 
             \large ABSTRACT \end{center}\bigskip 
      \begin{center}\begin{large}}{\end{large}\end{center} \end{quotation}}

\newenvironment{Presented}{\begin{quotation} \begin{center} 
             PRESENTED AT\end{center}\bigskip 
      \begin{center}\begin{large}}{\end{large}\end{center} \end{quotation}}

\input econfmacros.tex

\usepackage{color}

\newcommand\pubdate{\today}

\def\affiliation{
On behalf of the LHCb Experiment, \\
Department of Physics \\
University of Genova, Genova, Italy }

\begin{document}

\large
\begin{titlepage}
\pubblock

\vfill
\Title{Charmless B decays at LHCb}
\vfill

\Author{ Roberta Cardinale  }
\Address{\affiliation}
\vfill
\begin{Abstract}
The study of charmless $b$-hadron decays provides information for
testing the CKM picture of CP violation in the Standard
Model. In addition, as they can proceed through loop diagrams, they are
also sensitive to physics beyond the Standard Model. 
A review of recent results from LHCb on charmless $b$-hadron decays is
presented. 
\end{Abstract}
\vfill

\begin{Presented}
The Second Annual Conference\\
 on Large Hadron Collider Physics \\
Columbia University, New York, U.S.A \\ 
June 2-7, 2014
\end{Presented}
\vfill
\end{titlepage}
\def\thefootnote{\fnsymbol{footnote}}
\setcounter{footnote}{0}
%

\normalsize 


\section{Introduction}
Charmless $b$-hadron decays play a central role testing ground for the
Standard Model. Recent results using data collected in 2011
at the LHCb detector~\cite{Alves:2008zz} at
$\sqrt{s}=7\unitm{TeV}$, corresponding to an integrated luminosity of $\sim 1
\unitm{fb^{-1}}$, are presented.

\section{Search for $\Lambda_{b}^{0}(\Xi_{b}^{0}) \to K_{s}^{0} p h^{-}$}
The study of $b$-baryons decays is almost an unexplored field. Hadronic three-body $b$-baryons
decays to charmless final states, which have not been observed yet, can
provide the possibility to study hadronic decays and to search for CP
violation. In these proceedings are presented the branching fractions
measurements of beauty baryons decays to the
final states $K_{s}^{0}p\pi^{-}$ and $K_{s}^{0}pK^{-}$, determined relative to the $B^{0} \to K^{0}_{s}
\pi^{+} \pi^{-}$ decay used as normalisation channel~\cite{Aaij:2014lpa}.
Each $b$-hadron decay is reconstructed by combining two charged tracks
with a $K^{0}_{s}$ candidate. The $K^{0}_{s}$ candidates are
reconstructed in the $\pi^{+} \pi^{-}$ final states using two
different categories. The Long candidates have hits both in
the vertex detector and in the tracking stations downstream of the
dipole magnet while the Downstream candidates have not track
segments in the vertex detector but only in the tracking
stations. Events are triggered and selected in a similar way both for
the signal modes and the normalisation channel, exploiting the topology
of three-body decays and the $b$-hadron kinematic
properties. Intermediate states containing charmed hadrons are
excluded from the signal sample and studied separately.\\
\begin{figure}[htb]
\centering
\includegraphics[scale=0.4]{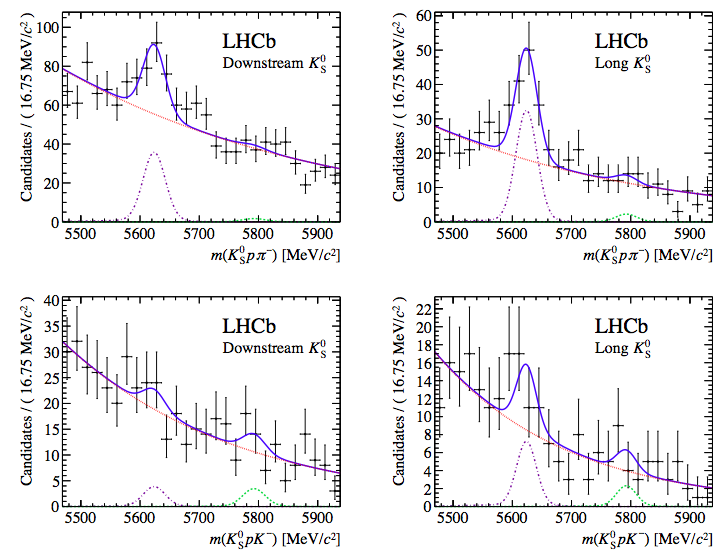}
\caption{Invariant mass distribution of (top) $K_{s}^{0}p\pi^{-}$ and
  (bottom) $K_{s}^{0} p K^{-}$ selected candidates for the (left) Downstream
  and (right) Long $K_{s}^{0}$ categories.}
\label{fig:lambda}
\end{figure}
The decay channel $\Lambda_{b}^{0} \to K_{s}^{0} p \pi^{-}$ is
observed for the first time with a significance level of
$8.6\sigma$ and its branching fraction is measured to be
$${\mathcal B}(\Lambda_{b}^{0} \to \bar K^{0} p \pi^{-})  = (1.26 \pm
0.19 \pm 0.09 \pm 0.34 \pm 0.05) \times 10^{-5},$$ where the
first uncertainty is statistical, the second systematic and the third
and the fourth related to the uncertainty on the ratio of
fragmentation fraction, $f_{\Lambda_{b}^{0}}/f_{d}$ and on the branching
fraction of the $B^{0} \to K^{0} \pi^{+} \pi^{-}$ decay respectively. 
The CP asymmetry integrated over the phase-space of the observed $\Lambda_{b}^{0} \to K_{s}^{0} p \pi^{-}$ decay is found to be
$${\mathcal A}^{CP}(\Lambda_{b}^{0} \to K_{s}^{0} p \pi^{-}) = 0.22 \pm
0.13 \, {\rm (stat)} \pm 0.03 \, {\rm (syst)}.$$
No significant signals are seen for the $\Lambda_{b}^{0} \to K^{0}_{s} p
K^{-}$ decay and for the $\Xi_{b}^{0}$ decays and upper limits on their
branching fractions are set to
\begin{align*}
{\mathcal B}(\Lambda_{b}^{0} \to \bar K^{0} p K^{-})  & < 3.5 (4.0)
\times 10^{-6} \; {\rm at} \;  90\% \; (95\%) \; {\rm CL}\\
f_{\Xi_{b}^{0}}/f_{d} \times {\mathcal B}(\Xi_{b}^{0} \to
\bar{K^{0}}p\pi^{-}) & <1.6 (1.8) \times 10^{-6} \;  {\rm at}  \;
90\% \; (95\%) \;{\rm  CL}\\
f_{\Xi_{b}^{0}}/f_{d} \times {\mathcal B}(\Xi_{b}^{0} \to
\bar{K^{0}}pK^{-}) &<1.1 (1.2) \times 10^{-6}\;  {\rm at}  \;  90\% \;
(95\%) \; {\rm CL}
\end{align*}

\section{Effective lifetime measurements of the $B_{s}^{0} \to K^{+}
  K^{-}$, $B^{0} \to K^{+} \pi^{-}$ and $B^{0}_{s} \to \pi^{+} K^{-}$ decays}
\begin{figure}[htb]
\centering
\includegraphics[scale=0.15]{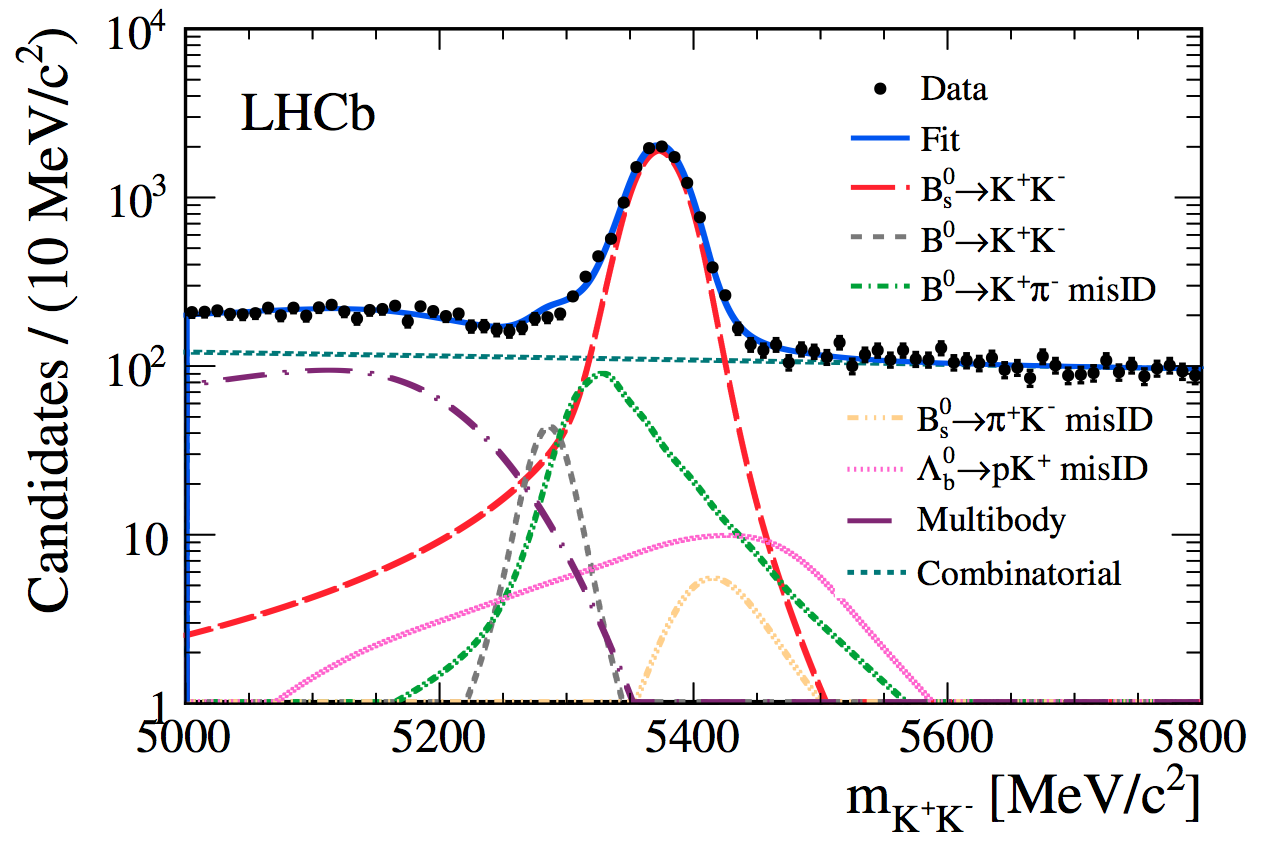}
\includegraphics[scale=0.167]{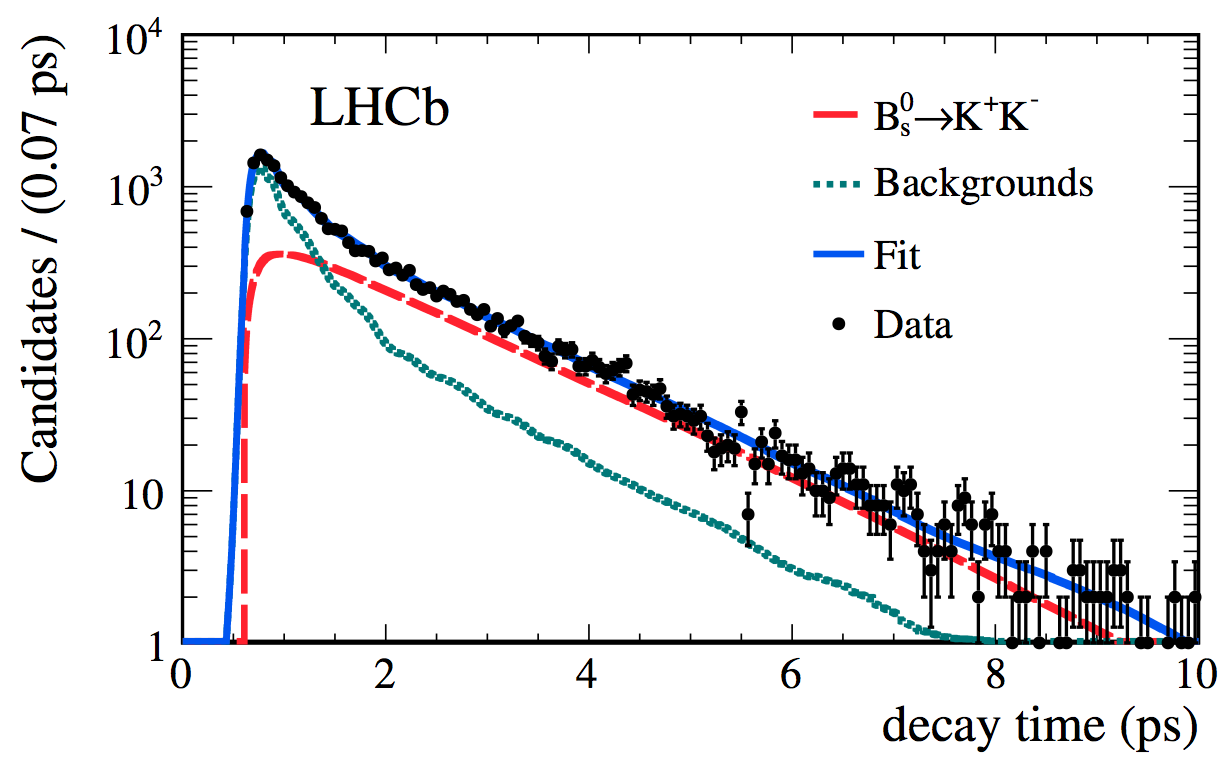}
\caption{Fit to the $KK$ invariant mass spectrum and to the
  reconstructed decay times.}
\label{fig:lifetime}
\end{figure}

The effective lifetime measurement of the $B_{s}^{0} \to K^{+}
  K^{-}$ decay, recently measured by LHCb with high precision,  is of
  great interest as it can constrain contributions
  from new physical phenomena to the $B^{0}_{s}$ system.
In addition the $B^{0} \to K^{+} \pi^{-}$ and
  $B^{0}_{s} \to K^{+} \pi^{-}$ lifetimes, which contribute to the world
  average of $\tau(B^{0})$ and $\tau(B^{0}_{s})$, are measured~\cite{Aaij:2014fia}.
The analysis uses a data driven approach to correct for the decay time
acceptance introduced by the trigger and the final selection. The
procedure consists in
extracting the per-event acceptance function directly from data. 
The
effective lifetimes are then determined using a factorised
fit to the mass and decay time distributions (see Figure~\ref{fig:lifetime}).
The measured $B_{s}^{0} \to K^{+} K^{-}$ lifetime is
$$ \tau_{B_{s}^{0} \to K^{+}
  K^{-}} = 1.407 \pm 0.016 \, {\rm (stat)} \pm 0.007 \, {\rm (syst)} \unitm{ps}$$
which is the world best measurement and is compatible with the SM
prediction. The dominant contribution to the systematic uncertainty
come from the contamination from misidentified $B \to h^{+} h^{'-}$
background channels. The measured lifetimes for $B^{0} \to K^{+} \pi^{-}$ and
$B^{0}_{s} \to \pi^{+} K^{-}$ decays are
\begin{align*}
\tau_{B^{0} \to K^{+} \pi^{-}} &= 1.524 \pm 0.011 \, {\rm (stat)} \pm
0.004 \,
{\rm (syst)} \unitm{ps}\\
\tau_{B^{0}_{s} \to \pi^{+} K^{-}} &= 1.60 \pm 0.06 \, {\rm (stat)} \pm 0.01
\, {\rm (syst)} \unitm{ps}
\end{align*}

\section{Measurement of CP violation in the phase space of $B^{\pm}
  \to K^{+} K^{-} \pi^{\pm}$ and $B^{\pm} \to \pi^{+} \pi^{-} \pi^{\pm}$}
\begin{figure}[htb]
\centering
\includegraphics[scale=0.4]{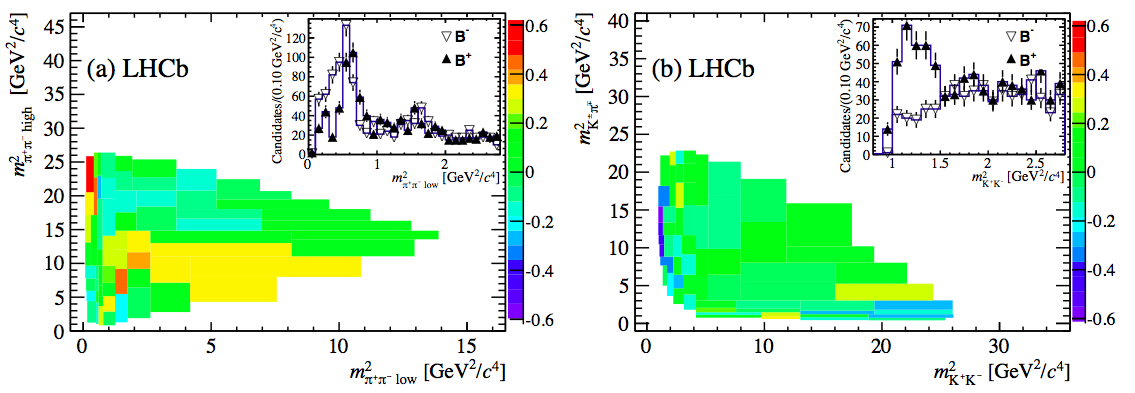}
\caption{Asymmetries of the number of events in bin of the Dalitz plot
for (a) $B^{\pm} \to
\pi^{\pm} \pi^{-} \pi^{+}$ and (b) $B^{\pm} \to \pi^{\pm} K^{+} K^{-}$. The
inset figures show the projections of the number of events in bins of
(a) $m^{2}_{\pi^{+}\pi^{-} \rm low}$ variable for $m^{2}_{\pi^{+} \pi^{-}
  \rm high}>15 \unitm{GeV/}c^{2}$ and (b) the $m^{2}_{K^{+}K^{-}}$ variable.}
\label{fig:3h}
\end{figure}

Charmless decays of $B$ mesons to three hadrons are dominated by
quasi-two body processes involving intermediate resonant states. The
rich interference pattern makes them favorable for the investigations
of CP asymmetries that are localized in the phase space. Interference
between intermediate states of the decay can introduce large strong
phase differences which can explain local asymmetries in the phase
space~\cite{Zhang:2013oqa, Bhattacharya:2013cvn}. Another explanation focuses on final-state $KK \leftrightarrow
\pi\pi$ rescattering, which can occur between decay channels with the
same flavour quantum numbers~\cite{Bhattacharya:2013cvn, Bediaga:2013ela}.
CP violation in the phase space of
$B^{+} \to K^{+} K^{-} \pi^{+}$ and $B^{+} \to \pi^{+} \pi^{-}
\pi^{+}$ is measured~\cite{Aaij:2013bla}.\\
Events are selected requiring that the three charged tracks satisfy
selection criteria related to their transverse momenta, vertex and
track quality. Final state kaons and pions are further selected.\\Raw asymmetries are extracted from an unbinned
maximum likelihood fit to the mass spectra of the selected
candidates and then corrected for detector induced effects
and for the $B^{\pm}$ meson production asymmetry
$$ A_{\rm CP} = A_{\rm raw} - A_{D}(\pi^{\pm}) - A_{P}(B^{\pm}) $$
The $\pi^{\pm}$ detection asymmetry, $A_{D}(\pi^{\pm})$, is calculated using the ratio of
full to partially reconstructed $D^{*+} \to \pi^{+} D^{0}$
decays~\cite{Aaij:2012cy}, while the production asymmetry, $A_{P}(B^{\pm}) $, is evaluated using $B^{\pm} \to J/\psi
K^{\pm}$ decay as control channel. The CP asymmetries are found to be
\begin{align*}
A_{CP} (B^{\pm} \to \pi^{\pm} K^{+} K^{-}) &= -0.141 \pm 0.040 \, {\rm (stat)} \pm
 0.018 {\rm (syst)}\pm  0.007 \, (A_{CP}(J/\psi K)) \\
A_{CP} (B^{\pm} \to \pi^{\pm} \pi^{+} \pi^{-}) &= 0.117 \pm 0.021 \,{\rm (stat)} \pm
0.009 \, {\rm (syst)}   \pm 0.007 \, (A_{CP}(J/\psi K))\\
\end{align*}
where the first uncertainty is statistical, the second is the systematic
uncertainty and the third is due to the uncertainty on the measurement
of the CP asymmetry of the $B^{\pm} \to J/\psi K^{\pm}$ decay. These
measurements represent the first
evidence of inclusive CP asymmetries of the $B^{\pm} \to K^{+} K^{-}
\pi^{\pm}$ and $B^{\pm} \to \pi^{+} \pi^{-}\pi^{\pm}$ decays with
significances of $3.2\sigma$ and $4.9\sigma$ respectively.\\Asymmetry distributions over the phase
space have been studied, as reported in Figure~\ref{fig:3h}, where the raw asymmetries in each bin of the
Dalitz plot are shown.\\ 
For the $B^{\pm} \to \pi^{\pm} K^{+} K^{-}$ decays a large negative charge
asymmetry is observed in the low \mbox{$m^{2}_{K^{+}K^{-}} < 1.5 \unitm{GeV^{2}/}c^{2}$} where no resonant contribution is expected. For $B^{\pm} \to
\pi^{\pm} \pi^{-} \pi^{+}$ decays, a large positive charge asymmetry is measured in the
low $m^{2}_{\pi^{+}\pi^{-} \rm low}<0.4 \unitm{GeV/}c^{2}$ and in the high
$m^{2}_{\pi^{+} \pi^{-}
  \rm high}>15 \unitm{GeV/}c^{2}$, not clearly associated to a
resonant state.  Unbinned extended maximum likelihood fits are performed
to the mass spectra of the candidates in the regions where large raw
asymmetries are found.
The local charge asymmetries for the two regions are measured to be
\begin{align*}
A^{\rm reg}_{\rm CP}(B^{\pm} \to K^{+} K^{-} \pi^{\pm}) &= -0.648 \pm
0.070 \,  {\rm (stat)}
\pm 0.013 \,  {\rm (syst)} \pm 0.007 \, (A_{CP}(J/\psi K))\\
A^{\rm reg}_{\rm CP}(B^{\pm} \to \pi^{+} \pi^{-} \pi^{\pm}) &= -0.584
\pm 0.082 \,  {\rm (stat)}
\pm 0.027 \, {\rm (syst)} \pm 0.007 \, (A_{CP}(J/\psi K))
\end{align*}
where the first uncertainty is statistical, the second is the systematic
uncertainty and the third is due to the uncertainty on the measurement
of the CP asymmetry of the $B^{\pm} \to J/\psi K^{\pm}$ decay.
Those results along with recent theoretical
developments, may indicate new mechanisms for CP
asymmetries~\cite{Zhang:2013oqa, Bhattacharya:2013cvn,
  Bediaga:2013ela, Xu:2013dta}.
\section{Measurement of polarization amplitudes and CP asymmetries in
  $B^{0} \to \phi K^{*0}$ decays}
In the Standard Model the $B^{0} \to \phi K^{*0}$ decay is expected to
proceed mainly via a gluonic
penguin diagram. For this reason the measurement of CP violation in
this decay is sensitive to possible physics beyond the Standard Model,
arising in the penguin
loop. Since this decay involves a spin-0 B meson decaying into two
spin-1 vector mesons, due to angular momentum conservation, there are
only three independent configurations of the final state spin vectors.
They can be written in term of a
longitudinal polarization, ${\mathcal A}_{0}$, and two transverse
components with collinear, ${\mathcal A}_{||}$, and orthogonal, ${\mathcal A}_{\perp}$,
polarizations.

Angular analyses have shown that the
longitudinal and transverse components in this decay have roughly
equal amplitudes. Similar results have been observed also in other $B
\to VV$ transitions in contrast to tree-level
decays~\cite{delAmoSanchez:2010mz, Abe:2004mq, Aubert:2006fs, Aaij:2011aj}.
The different
behaviour of tree and penguin decays has attracted much theoretical
attention~\cite{Kagan:2004uw, Datta:2007qb}. In addition to the P-wave
amplitudes, there are also contributions where $K^{+}K^{-}$ or
$K^{+}\pi^{-}$ are produced in a spin-0 (S-wave) state,
($A_{s}^{K^{+}K^{-}}$ and $A_{s}^{K^{+}\pi^{-}}$).

Polarization amplitudes
and phases are measured by LHCb performing the studies of the angular
distributions of the decay products~\cite{Aaij:2014tpa}.
Candidates are selected from charged tracks with high transverse
momentum and impact parameter. Pions and kaons are then selected using
particle identification information provided by the RICH
detectors. The resulting charged tracks are
combined to form $\phi$ and $K^{*0}$ meson candidates requiring the
invariant mass to be close to the known mass. Kinematic and
topological variables are then used in a geometric likelihood method
to further suppress background, obtaining about 1800 candidates. 
A simultaneous fit to the invariant masses and angular
observables distributions is performed. The
angular analysis results are reported in Table~\ref{tab:angularfit}.
\begin{table}[htbp]
\begin{center}
\begin{tabular}{ccc}  
\hline
Parameter & Definition &  Fitted value\\
\hline
$f_{\rm L}$ & $0.5(|A_{0}|^2/F_{P} +|\bar A_{0}|^2/\bar F_{P}  )$  &$0.497 \pm 0.019 \pm 0.015$\\
$f_{\perp}$ & $0.5(|A_{\perp}|^2/F_{P} +|\bar A_{\perp}|^2/\bar F_{P}  )$ &$0.221 \pm 0.016  \pm 0.013$\\
$f_{s}(K\pi)$ & $0.5(|A^{K\pi}_{s}|^2+|\bar A^{K\pi}_{s}|^2)$ & $0.143 \pm 0.013 \pm 0.012$\\
$f_{s}(KK)$ & $0.5(|A^{KK}_{s}|^2+|\bar A^{KK}_{s}|^2)$&$0.122 \pm 0.013\pm 0.008$\\
$\delta_{\perp}$ & $0.5( \rm arg \, A_{\perp} + \rm arg \, \bar A_{\perp} )$ & $2.633 \pm 0.062 \pm 0.037$\\
$\delta_{\parallel}$ & $0.5( \rm arg \, A_{||} + \rm arg \, \bar A_{||} )$&$2.562 \pm 0.069 \pm 0.040$\\ 
$\delta_{s}(K\pi)$ &$0.5( \rm arg \, A^{K\pi}_{s}+ \rm arg \, \bar A^{K\pi}_{s})$ &$2.222 \pm 0.063 \pm 0.081$\\ 
$\delta_{s}(KK)$ &$0.5( \rm arg \,A^{KK}_{s}+ \rm arg \, \bar A^{KK}_{s})$  &$2.481 \pm 0.072 \pm 0.048$\\ 
$A_{0}^{\rm CP}$ &$(|A_{0}|^2/F_{P} -|\bar A_{0}|^2/\bar F_{P}  )/(|A_{0}|^2/F_{P} +|\bar A_{0}|^2/\bar F_{P}  )$ &$-0.003 \pm 0.038\pm 0.005$\\
$A_{\perp}^{\rm CP}$ & $(|A_{\perp}|^2/F_{P} -|\bar A_{\perp}|^2/\bar F_{P}  )/(|A_{\perp}|^2/F_{P} +|\bar A_{\perp}|^2/\bar F_{P}  )$ &$+0.047 \pm 0.072\pm 0.009$\\
$A_{s}(K\pi)^{\rm CP}$ & $(|A^{K\pi}_{s}|^2-|\bar A^{K\pi}_{s}|^2)/(|A^{K\pi}_{s}|^2+|\bar A^{K\pi}_{s}|^2)$&$+0.073 \pm 0.091\pm 0.035$\\
$A_{s}(KK)^{\rm CP}$ & $(|A^{KK}_{s}|^2-|\bar A^{KK}_{s}|^2)/(|A^{KK}_{s}|^2+|\bar A^{KK}_{s}|^2)$ &$-0.209 \pm 0.105\pm 0.012$\\
$\delta_{\perp}^{\rm CP}$ &$0.5( \rm arg \, A_{\perp} - \rm arg \, \bar A_{\perp} )$  &$+0.062 \pm 0.062 \pm 0.006$\\
$\delta_{\parallel}^{\rm CP}$ & $0.5( \rm arg \, A_{||} - \rm arg \, \bar A_{||} )$&$+0.045 \pm 0.069 \pm 0.015$\\
$\delta_{s}(K\pi)^{\rm CP}$ &$0.5( \rm arg \, A^{K\pi}_{s}- \rm arg \, \bar A^{K\pi}_{s})$ &$0.062 \pm 0.062 \pm 0.022$\\ 
$\delta_{s}(KK)^{\rm CP}$ & $0.5( \rm arg \,A^{KK}_{s}- \rm arg \, \bar A^{KK}_{s})$&$0.022 \pm 0.072 \pm 0.004$\\ 
\hline
\end{tabular}
\end{center}
\caption{Parameters measured in the angular analysis. The first and
  second uncertainties are statistical and systematic,
  respectively. The P- and S-wave fractions are defined as $F_{P} =
  |A_{0}|^{2} + |A_{||}|^{2} + |A_{\perp}|^{2}$, $F_{P} =
  |A^{K\pi}_{s}|^{2} + |A^{KK}_{s}|^{2}$, $F_{P} + F_{s} =1$.}
\label{tab:angularfit}
\end{table}
The P-wave parameters
are consistent with, but more precise than previous measurements and
the value of $f_{\rm L}$ indicates that longitudinal and transverse polarizations have similar
size~\cite{Aubert:2008zza, Prim:2013nmy}. Significant S-wave contributions, $A_{s}^{K^{+}K^{-}}$ and $A_{s}^{K^{+}\pi^{-}}$, are found in both the $K^{+}\pi^{-}$
and $K^{+}K^{-}$ systems. The CP asymmetries in both the amplitudes
and the phases are consistent with zero. The largest systematic
uncertainty on the angular analysis is due to the understanding of the
detector acceptance which is determined from simulated events.\\
\section{Conclusions}
An overview of the latest LHCb results on charmless $b$-hadron decays
has been given. First observation of $b$-baryons decays to hadronic
three-body charmless final states has been obtained. The measured effective
lifetime in the $B_{s}^{0} \rightarrow K^{+}K^{-}$ decay has been
found compatible with the SM expectation. In the $B^{\pm}
  \to K^{+} K^{-} \pi^{\pm}$ and $B^{\pm} \to \pi^{+} \pi^{-}
  \pi^{\pm}$ decays, a large CP asymmetry has been found in regions of
  the Dalitz which do not correspond to resonant contributions.
  This may indicate new mechanisms for CP
asymmetries. More interesting results are expected using the complete 2011
and 2012 available data samples which correspond to an integrated
luminosity of $\sim 3 \unitm{fb^{-1}}$. 

\end{document}

%% file: econfmacros.tex
\def\beq{\begin{equation}}
\def\eeq#1{\label{#1}\end{equation}}
\def\eeqn{\end{equation}}

\def\beqa{\begin{eqnarray}}
\def\eeqa#1{\label{#1}\end{eqnarray}}
\def\eeqan{\end{eqnarray}}

\let\bar=\overbar

\def\Dslash{\not{\hbox{\kern-4pt $D$}}}
\def\dslash{\not{\hbox{\kern-2pt $\del$}}}

\def\msb{{\bar{\ssstyle M \kern -1pt S}}}

